\documentclass[useAMS]{mn2e}

\usepackage[dvips]{graphicx}
\usepackage{amssymb}
\usepackage{amsmath}
\def\jnl@style{\rm}
\def\aaref@jnl#1{{\jnl@style#1}}

\def\aaref@jnl#1{{\jnl@style#1\thinspace}}

\def\aj{\aaref@jnl{AJ}}                   
\def\araa{\aaref@jnl{ARA\&A}}             
\def\apj{\aaref@jnl{ApJ}}                 
\def\apjl{\aaref@jnl{ApJ}}                
\def\apjs{\aaref@jnl{ApJS}}               
\def\ao{\aaref@jnl{Appl.~Opt.}}           
\def\apss{\aaref@jnl{Ap\&SS}}             
\def\aap{\aaref@jnl{A\&A}}                
\def\aapr{\aaref@jnl{A\&A~Rev.}}          
\def\aaps{\aaref@jnl{A\&AS}}              
\def\azh{\aaref@jnl{AZh}}                 
\def\baas{\aaref@jnl{BAAS}}               
\def\jrasc{\aaref@jnl{JRASC}}             
\def\memras{\aaref@jnl{MmRAS}}            
\def\mnras{\aaref@jnl{MNRAS}}             
\def\pra{\aaref@jnl{Phys.~Rev.~A}}        
\def\prb{\aaref@jnl{Phys.~Rev.~B}}        
\def\prc{\aaref@jnl{Phys.~Rev.~C}}        
\def\prd{\aaref@jnl{Phys.~Rev.~D}}        
\def\pre{\aaref@jnl{Phys.~Rev.~E}}        
\def\prl{\aaref@jnl{Phys.~Rev.~Lett.}}    
\def\pasp{\aaref@jnl{PASP}}               
\def\pasj{\aaref@jnl{PASJ}}               
\def\qjras{\aaref@jnl{QJRAS}}             
\def\skytel{\aaref@jnl{S\&T}}             
\def\solphys{\aaref@jnl{Sol.~Phys.}}      
\def\sovast{\aaref@jnl{Soviet~Ast.}}      
\def\ssr{\aaref@jnl{Space~Sci.~Rev.}}     
\def\zap{\aaref@jnl{ZAp}}                 
\def\nat{\aaref@jnl{Nature}}              
\def\iaucirc{\aaref@jnl{IAU~Circ.}}       
\def\aplett{\aaref@jnl{Astrophys.~Lett.}} 
\def\apspr{\aaref@jnl{Astrophys.~Space~Phys.~Res.}}
\def\bain{\aaref@jnl{Bull.~Astron.~Inst.~Netherlands}} 
\def\fcp{\aaref@jnl{Fund.~Cosmic~Phys.}}  
\def\gca{\aaref@jnl{Geochim.~Cosmochim.~Acta}}   
\def\grl{\aaref@jnl{Geophys.~Res.~Lett.}} 
\def\jcp{\aaref@jnl{J.~Chem.~Phys.}}      
\def\jgr{\aaref@jnl{J.~Geophys.~Res.}}    
\def\jqsrt{\aaref@jnl{J.~Quant.~Spec.~Radiat.~Transf.}}
\def\memsai{\aaref@jnl{Mem.~Soc.~Astron.~Italiana}}
\def\nphysa{\aaref@jnl{Nucl.~Phys.~A}}   
\def\physrep{\aaref@jnl{Phys.~Rep.}}   
\def\physscr{\aaref@jnl{Phys.~Scr}}   
\def\planss{\aaref@jnl{Planet.~Space~Sci.}}   
\def\procspie{\aaref@jnl{Proc.~SPIE}}   

\begin{document}

\title[VW\,Hyi: optical spectroscopy and Doppler tomography]
  {VW\,Hyi: optical spectroscopy and Doppler tomography}
\author[A. J. Smith et al.]
  {Amanda J. Smith$^{1,}\thanks{E-mail: amanda.smith@open.ac.uk}$, Carole A. Haswell$^{1,}$\thanks{Visiting astronomer, Cerro Tololo Inter-American Observatory, National
Optical Astronomy Observatories, which are operated by the Association of
Universities for Research in Astronomy, under contract with the National
Science Foundation.}, Robert I. Hynes$^{2,} \dagger$\\
$^1$Department of Physics \& Astronomy, The Open University, Walton Hall, Milton Keynes, MK7 6AA, UK\\
$^2$Department of Physics \& Astronomy, Louisiana State University, 202 Nicholson Hall, Tower Drive, Baton Rouge, \\ Louisiana 70803-4001, USA}
\date{Accepted. Received}
\pagerange{\pageref{firstpage}--\pageref{lastpage}}
\pubyear{2006}
\maketitle \label{firstpage}
\begin{abstract}
We present high quality optical spectroscopy of the SU\,UMa-subtype dwarf nova, VW\,Hyi taken while the system was in quiescence. An S-wave is executed by the emission cores of the Hydrogen Balmer lines and by the emission lines of He\,{\sc i}, Ca\,{\sc ii}, Fe\,{\sc ii} and He\,{\sc ii}. Using Doppler tomography we show it originates in the accretion stream-disc impact region.
The He\,{\sc ii} emission is strongly phase-dependent, suggesting it originates exclusively within a hot cavity at the initial impact. We map the ionization structure of the stream-disc interaction region. One possible interpretation of this is that the Balmer hotspot lies downstream of the He\,{\sc ii} hotspot in the outer accretion disc, with the He\,{\sc i} and Ca\,{\sc ii} hotspots at intermediate locations between the two. This suggests that Balmer emission is suppressed until material has cooled somewhat downstream of the impact site and is able to recombine. We favour a phase offset of $0.15 \pm 0.04$ between the
photometric ephemeris and inferior conjunction of the mass donor. The white dwarf contributes significantly to the optical continuum, with broad Balmer absorption and narrow Mg\,{\sc ii}\,$\lambda4481$ absorption clearly apparent. This latter feature yields the gravitational redshift: $v_{\rm grav}=38\pm21\,{\rm km\,s}^{-1}$, so $M_{1}=0.71^{+0.18}_{-0.26}\,{\rm M_{\odot}}$. This implies $M_2= 0.11 \pm 0.03\,{\rm M_\odot}$ and hence the donor is not a brown dwarf. A prominent Balmer jump is also observed. We note that the previously accepted system parameters for both VW\,Hyi and WX\,Hyi incoporate an algebraic error, and we provide a recalculated $M_{1}(q)$ plane for WX\,Hyi.

\end{abstract}

\begin{keywords}
accretion, accretion discs - techniques: spectroscopic - stars: dwarf novae - stars: individual: VW\,Hyi - novae, cataclysmic variables
\end{keywords}

\section{Introduction}

The close interacting binary, VW\,Hyi, is a cataclysmic variable (CV), a system in which a Roche-lobe filling low-mass, late-type main sequence donor star transfers matter onto a white dwarf (WD) primary via an accretion disc.
With a quiescent visual magnitude of 13.8, VW\,Hyi is the brightest example of the SU\,UMa-type dwarf novae (DNe). These systems display two distinct modes of outburst, namely, normal DNe outbursts and less frequent larger amplitude superoutbursts. These two modes of outburst occur in VW\,Hyi typically about every 30\,d and 180\,d reaching a peak visual magnitude of 9.5 and 8.7 respectively.

The DNe outburst can be understood in the context of the disc instability model (DIM) \nocite{O74}({Osaki} 1974). In this model a thermal-viscous limit-cycle instability \nocite{MM-H81}({Meyer} \& {Meyer-Hofmeister} 1981), caused by a sensitive temperature dependence of opacity in regions of partial ionization in the disc, induces a switch between low and high viscosity states, and consequently between low and high mass transfer states. There remain some questions about the appropriateness of this model however \nocite{L01}(see {Lasota} 2001, for a review). The superoutburst, distinguishable by a photometric modulation known as superhumps, can be explained by a model combining the thermal instability of the DNe outbursts with a tidal instability caused by the action of the donor in a 3:1 resonance with particle orbits in the accretion disc. This is the thermal-tidal instability (TTI) model of \nocite{O89,O96}{Osaki} (1989, 1996).

As with other SU\,UMa-type systems, VW\,Hyi has a period that lies below the CV period gap. Its period, $P_{\rm orb}=1.78\,{\rm h}$, was determined photometrically by way of the orbital hump observed at quiescence which arises from the phase dependent visibility of the accretion stream-disc impact region (i.e. the hotspot) throughout the orbit \nocite{V74,vADG87}({Vogt} 1974; {van Amerongen} {et~al.} 1987).

\subsection{The hotspot}

Hotspot emission should arise primarily from dissipation of kinetic energy as the in-falling stream impacts the disc edge, merges with the Keplerian motion of the disc and radiates away during the encounter.
The details of the stream-disc impact region are as yet poorly understood. Study of the interaction between the two flows, the stream from $L_{1}$ and the disc itself, may lead to a better understanding of the viscosity in accretion discs.

The flows are highly supersonic, and so the picture of the hydrodynamics of the impact point is believed to be one which involves the development of shock fronts: one in the disc material and one in the incoming stream \nocite{RS-C87}({Rozyczka} \& {Schwarzenberg-Czerny} 1987). Emission may represent a mixture of accretion stream and disc material compressed between these shock surfaces \nocite{MHS90}({Marsh} {et~al.} 1990), or emission may be seen at two distinct velocities. The smoothed particle hydrodynamics (SPH) simulations of \nocite{FHM04}{Foulkes} {et~al.} (2004) show two distinct S-waves in simulated trailed spectrograms.

Other numerical investigations of the stream-disc impact have been carried out by \nocite{AL96,AL98}{Armitage} \& {Livio} (1996, 1998), and by \nocite{KSH01}{Kunze}, {Speith} \& {Hessman} (2001). The latter paper shows, through high resolution SPH simulations, a substantial vertical deflection of the impacting accretion stream after undergoing shock interaction at the hotspot. The material is also seen to overflow the disc, re-impacting the disc surface at a secondary hotspot close to the circularization radius at orbital phase 0.5.

\subsection{System parameters}

\begin{figure}
\centering
\includegraphics[width=1.04\columnwidth]{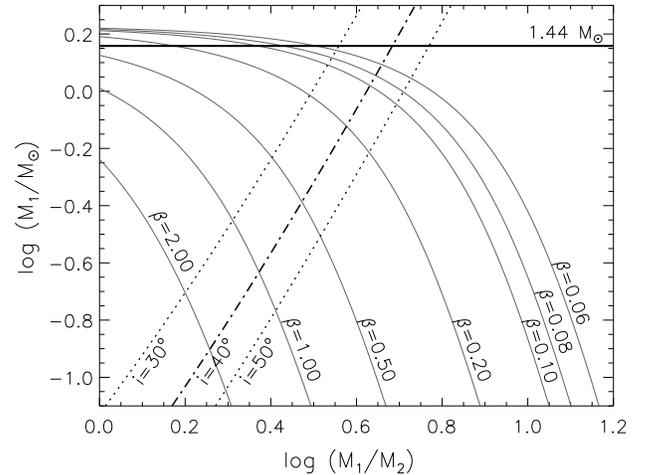}
\caption{$M_{1}(q)$ plane for WX\,Hyi after figure 7 of Schoembs \& Vogt (1981) for a range of values of their $\beta$ parameter and using their value for $K_{1}$ of $67\pm6\,{\rm km\,s}^{-1}$. Broken lines indicate the range of values of orbital inclination considered by Schoembs \& Vogt. In order to directly compare with Schoembs \& Vogt's figure, the abscissa label we use here follows their definition of $q$, i.e. $M_{1}/M_{2}$, this being the reciprocal of the definition we have adopted in the rest of this work ($M_{2}/M_{1}$).}
\label{SV}
\end{figure}

Despite being an extensively studied system, particularly in the UV, the basic parameters of VW\,Hyi remain uncertain. \nocite{SV81}{Schoembs} \& {Vogt} (1981) put forward an orbital inclination, $i$, of $60^{\circ}\pm10^{\circ}$, based on the fact that an orbital hump is observed in the lightcurve but no eclipse.
\nocite{SV81}{Schoembs} \& {Vogt} (1981) also determined component masses by a method based upon that of \nocite{W73}{Warner} (1973) relating the radial velocity semi-amplitude, $K_{1}$, of the emission lines and the projected rotation velocity of the disc $V_{\rm d}\,{\rm sin}i$ to the mass ratio, $q=M_{2}/M_{1}$. The estimation of $K_{1}$ using the emission lines in CVs has proved, in general, to be unreliable (see e.g. \nocite{HWS91}{Horne}, {Wood} \& {Stiening} (1991) for comparison between the spectroscopic determination of $K_{1}$ and the prediction of $K_{1}$ based on the photometric solution in the eclipsing system HT Cas). Often the hotspot distorts the apparent radial velocities as measured from the emission lines, even if only the line wings are used, as discussed by \nocite{M98}{Marsh} (1998). \nocite{SV81}{Schoembs} \& {Vogt} (1981) report $K_{1}=78\pm14\,{\rm km\,s}^{-1}$ from their radial velocity curve of $H\alpha$. Their tabulated system parameters are, however, inconsistent with this value: $M_{1}=0.63\,{\rm M_{\odot}}$, $q=1/6$ and $i=60^{\circ}$ give $K_{1}=57\,{\rm km\,s}^{-1}$. A missing exponent (4/3) on the $(1+q)$ term in their equation 7 appears to be the source of this discrepancy. It reappears in their subsequent equations, but has been omitted when calculating the orbital inclination in their figures 6 and 7. Based on the authors' method we have recalculated $M_{1}(q)$ for their values of $K_{1}$. For WX\,Hyi we show this in Figure \ref{SV}.
Taking the authors' assumptions at face value we find $M_{1}=0.92^{+0.18}_{-0.15}$ and $q=0.21^{+0.03}_{-0.02}$ for VW\,Hyi, and $M_{1}=1.14^{+0.26}_{-0.24}$ and $q=0.23^{+0.07}_{-0.04}$ for WX\,Hyi. However, we feel that these values are highly insecure, as the value of $\beta$ particularly may be quite different from that assumed, where $\beta$ is approximately the ratio of the primary radius to innermost accretion disc radius contributing to the line wings.

For VW\,Hyi, perhaps more credible, by measurement of the gravitational redshift from the shift in photospheric absorption lines of the white dwarf, \nocite{SCS97}{Sion} {et~al.} (1997) found $M_{1}=0.86^{+0.18}_{-0.32}\,{\rm M_{\odot}}$, the weakest part of this work being the estimate of the systemic velocity.
\nocite{WW02b}{Warner} \& {Woudt} (2002) assume that the minimum period observed for dwarf novae oscillations (DNOs) in VW\,Hyi reflects the Kepler period at the surface of the white dwarf, $P_{\rm K}$. Together with the mass-radius relationship for white dwarfs \nocite{N72}({Nauenberg} 1972), they calculate a value for $M_{1}$ of $0.702\,{\rm M_{\odot}}$. It seems appropriate to consider this value as a possible lower limit on $M_{1}$ as the observed minimum DNO period could exceed $P_{\rm K}$.
Most recently, from fitting of the J band spectral energy distribution of the donor, \nocite{MDT04}{Mennickent}, {Diaz} \&  {Tappert} (2004) suggest it to be of spectral type L0, putting $M_{2}\sim0.06\,{\rm M_{\odot}}$, and therefore implying a star below the hydrogen-burning minimum mass limit, a brown dwarf. If reliable, this would be the first detection of the secondary star.

There exists a relation between the fractional superhump excess, $\epsilon$, defined as $(P_{\rm sh}-P_{\rm orb})/P_{\rm orb}$ and the mass ratio $q$, where $P_{\rm sh}$ and $P_{\rm orb}$ are the superhump and orbital periods respectively. This has been calibrated empirically by \nocite{PKH05}{Patterson} {et~al.} (2005). For VW\,Hyi, $\epsilon=0.0331\pm0.0008$ \nocite{P98}({Patterson} 1998), and this indicates $q=0.148\pm0.004$. In this work we will assume $M_{1}=0.71^{+0.18}_{-0.26}\,{\rm M_{\odot}}$ (see Section \ref{grav}), and the value of $q$ derived from the superhump excess, $q=0.148\pm0.004$, unless stated otherwise.

In this paper, we present high quality optical spectroscopy of VW\,Hyi. The observations and data reduction are described in Section 2. In Section 3 we present our results, including Doppler maps of  VW\,Hyi and determination of the gravitational redshift of the white dwarf. Interpretation of these results is given in Section 4, which includes discussion of the ionization structure of the stream-disc impact and further discussion of the system parameters for VW\,Hyi. A summary is given in Section 5.

\section{Observations and data reduction}
We observed VW\,Hyi on 2003 February 14 01:09\,--\,02:56 UT using the CTIO 4.0\,m R-C spectrograph as detailed in Table \ref{tb:obs}. The AAVSO lightcurve for VW\,Hyi shows an outburst some 27 days before our observations, peaking on 2003 January 18. The timing of the subsequent outburst is unclear. An uncertain magnitude of 9.5 was recorded on 2003 February 16, while the following day the magnitude was confirmed to be 10.6, roughly three magnitudes above quiescence. This suggests that our observations took place just prior to the onset of outburst. The previous superoutburst took place in November 2002 \nocite{AAVSO}({Waagen} 2004).

During the observations the seeing was $1.4^{\prime\prime}$.
Using the KPGL1 grating and a Loral 3K CCD in conjunction with a $1.3^{\prime\prime}$ slit, a wavelength coverage of 3590\,--\,6686\,\AA\,and a FWHM resolution of $\sim3$\,\AA\,as measured from arc lines was obtained. We obtained 34 useful spectra, each with an integration time of 120\,s, together with three approximately equally interspersed HeNeAr arc-lamp exposures, giving $\sim85$ per cent orbital coverage. The spectra have a signal-to-noise ratio of $\sim50$. For the purposes of flux calibration, the standard star EG21 was observed.

The observations were reduced following standard procedures using {\sc IRAF} \footnote{{\sc IRAF} is distributed by the National Optical Astronomy Observatories, which are operated by the Association of Universities for Research in Astronomy, Inc., under cooperative agreement with the National Science Foundation.}.
The frames were first bias-subtracted, flat-fielded and sky-subtracted. The spectra were then extracted according to the optimal algorithm of \nocite{H86}{Horne} (1986). The arcs were extracted at the position of the corresponding object spectrum which were then used to wavelength calibrate the data by interpolation between neighbouring arcs. A zero-point correction of $+4.5\pm1.7\,{\rm km\,s}^{-1}$ was applied to the wavelength scale, derived from the O\,{\sc i}\,$\lambda5577$ sky emission line. The spectra were subsequently flux calibrated. All spectra were transformed to the heliocentric restframe.

Without correcting for slit losses, we convolved the flux calibrated average spectrum with the filter bandpass, hence deducing $V\lesssim14.5$. This indicates the system was quiescent. The system may have been fainter than the usual quiescent magnitude, but slit losses of 0.7 magnitudes are possible.

\begin{table}
\caption{2003 February 14, CTIO observations of VW\,Hyi}
\label{tb:obs}
\begin{center}
\begin{tabular}{|l|l|r|r|r|}

\hline
HJD-2452680 & Exposure & Exposure & N & Orbital \\
(mid-exposure) & & Time/s & & Phase\\
\hline

4.5383 & EG21 & 5.0 & 1 &  \\
4.5404 - 4.5416 & EG21 & 20.0 & 3 & \\
4.5589 - 4.5831 & VW\,Hyi & 120.0 & 14 & 0.677-0.003\\
4.5887 - 4.6220 & VW\,Hyi & 120.0 & 20 & 0.079-0.528\\

\hline

\end{tabular}
\end{center}
\end{table}

\section{Results}

\begin{figure*}
\centering
\includegraphics[width=0.72\textwidth,angle=90]{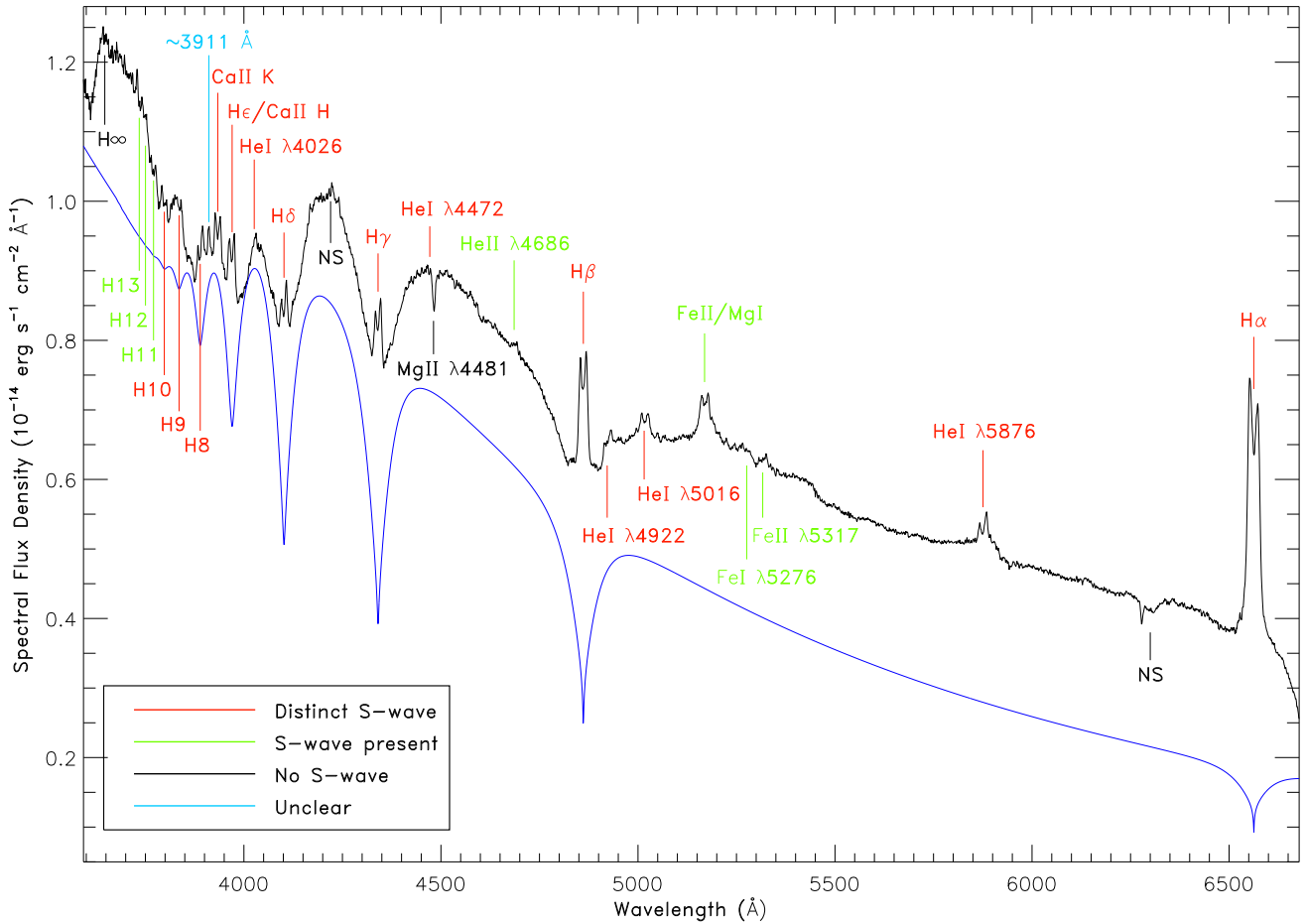}
\caption{Average spectrum of VW\,Hyi, in which each individual spectrum is given equal weighting and no correction for orbital motion has been made. Overplotted is a model white dwarf spectrum (solid blue line) (see Section \ref{swave}).
  }
\label{average}
\end{figure*}

The average of all 34 spectra is presented in Figure \ref{average}, and is quite remarkable in its detail and wealth of features.
The spectrum shows strong Balmer lines and a prominent Balmer jump. He\,{\sc i}, Fe\,{\sc ii} and Ca\,{\sc ii} emission features are also present, and He\,{\sc ii} emission is weak but clearly detected. H$\alpha$ is dominated by double-peaked emission: a typical signature of disc emission, due to the Doppler shifting of orbiting disc material.

The emission cores of the other lines in the Balmer series are flanked by broad absorption wings. We attribute this broad absorption to Stark broadening in the photosphere of the white dwarf, as the line width of this absorption would seem to preclude the possibility of kinematic broadening in the disc.
A conservative estimate for the base-to-base width of the H$\gamma$ absorption is $210$\,\AA. If this is due to Doppler broadening, a radial velocity $v_{\rm r}\simeq7300\,{\rm km\,s}^{-1}$ is required. Assuming $i=60^{\circ}$, the corresponding true velocity is $v_{\rm true}\simeq8400\,{\rm km\,s}^{-1}$. At the inner disc radius, the Keplerian velocity of disc material is at most $\sim 4200\,{\rm km\,s}^{-1}$, assuming the highest estimate for $M_{1}$, the inner disc radius to coincide with the white dwarf surface, and the mass-radius relation for white dwarfs of \nocite{N72}{Nauenberg} (1972).
In CVs, generally the accretion flow dominates at optical wavelengths, but it is not too unusual to see optical absorption features from the white dwarf as we do here, in systems where the mass-accretion rate is low for some reason \nocite{MHS87,HKS89}(e.g. {Marsh}, {Horne} \& {Shipman} 1987; {Hessman} {et~al.} 1989). These quiescent broad absorption features are seen in one previous spectral study of VW\,Hyi \nocite{HPS-C83}({Hassall} {et~al.} 1983), but are in contrast to the results of other studies also conducted in quiescence, where no such features are observed \nocite{SV81,Say89}({Schoembs} \& {Vogt} 1981; {Saygac} 1989).  This might suggest the system being in an unusually low state. However there is no clear indication of this from the AAVSO lightcurve.  In VW\,Hyi, the white dwarf is clearly revealed in quiescence in the UV \nocite{MS84,SHS95}({Mateo} \& {Szkody} 1984; {Sion} {et~al.} 1995).
The Balmer jump shows absorption in the Balmer continuum (i.e. $\lambda\,<\,3646$\,\AA). We refer to \nocite{GBT97}{G{\"a}nsicke}, {Beuermann} \&  {Thomas} (1997)'s spectrum of EK TrA, which shows a similar feature although less pronounced than that seen in our data.

\begin{figure*}
\centering
\includegraphics[height=9.7cm]{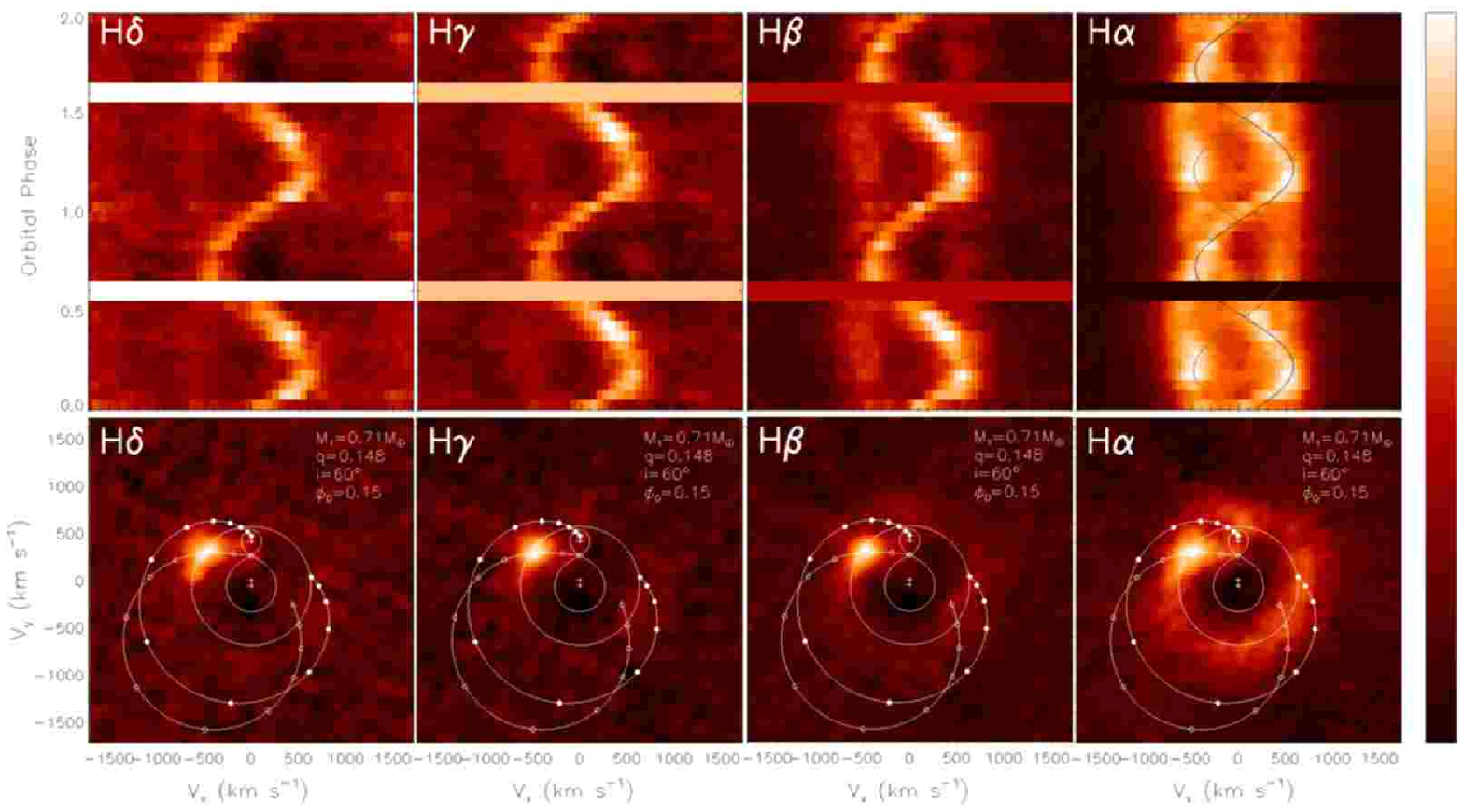}
\caption{Top: Trailed spectra of H$\alpha$, H$\beta$, H$\gamma$ and H$\delta$ emission respectively, where the spectra have been averaged into 20 orbital phase bins and the full cycle repeated for clarity. Overplotted on H$\alpha$ are sine waves tracing the motion of the main S-wave emission (black), and indicating the motion of a second possible isolated emission site (dark orange).
Bottom: Doppler maps of Balmer emission. The system parameters used in the construction of the maps are shown in the legends. The value for $i$ given by Schoembs \& Vogt (1981) was assumed. $\phi_{0}$ gives the orbital phase with respect to the photometric ephemeris of van Amerongen et al. (1987). The ballistic trajectory of the gas stream is represented by the solid line, with open circles denoting steps of 0.2$R_{\rm L1}$ along the stream. The `Kepler shadow', which shows the Keplerian velocity at equivalent positions along the ballistic trajectory, is represented by the dotted line with filled circles. The crosses, reading from top to bottom, denote the centre of mass of the secondary, the centre of mass of the system, and the centre of mass of the primary respectively. The Roche lobe of the primary is given by the dashed line, and the Roche lobe of the secondary is shown by a solid line. The circular solid line represents Keplerian velocity at the tidal truncation radius of the accretion disc.}
\label{trail}
\end{figure*}

The double-peaked emission component, clear in H$\alpha$, becomes progressively weaker towards the higher order Balmer lines, as seen in the trailed spectrograms in the top panels of Figure \ref{trail}. Here, the spectra have been normalized using a low-order spline fit to the continuum. We also see the S-waves executed by the emission cores, due to the periodic redshifting and blueshifting of an isolated emission component orbiting in the binary system. The amplitude of the S-wave is $\sim 500\,{\rm km\,s}^{-1}$. The emission appears to be strongest in the transition from red to blue. Away from the main S-wave, the double-peaked component of the emission is still variable. The enhanced regions of this double-peaked emission may be independent, but their phasing is roughly consistent with the sinusoidal motion of a single emission site in the disc, as indicated by the second overplotted sine wave on the trailed spectrum of H$\alpha$.

S-waves are clearly detected in the Balmer series almost right up to the Balmer limit \nocite{Smith}({Smith} 2006). The majority of the emission lines in the other species also show evidence of S-wave emission (see Figure \ref{average} and Sections \ref{hetxt} and \ref{caiitxt}).

At the suggestion of the referee, and for the purposes of comparison with previous studies, we employed the double Gaussian convolution method \nocite{SY80,S83}({Schneider} \& {Young} 1980; {Shafter} 1983) to examine the behaviour of the H$\alpha$ emission profile. Results of this are to be included in \nocite{Smith}{Smith} (2006). The value of $K_{\rm em}$ so obtained is in agreement with that expected from our chosen system parameters, but there is a strong likelihood of hotspot contamination even if large Gaussian separations are used, particularly here where the S-wave clearly dominates the disc emission (Figure \ref{trail}).

\subsection{Doppler tomography}
The indirect imaging technique of Doppler tomography \nocite{MH88}({Marsh} \& {Horne} 1988) translates orbital phase-resolved emission line profiles into a distribution of line emission in two-dimensional velocity space.
In this work, Doppler maps were produced using Tom Marsh's {\sc MOLLY} and {\sc DOPPLER} software. With the exception of Figure \ref{mem}, the Fourier-filtered back-projection (FFBP) method was used \nocite{M01}(see {Marsh} 2001, for a recent review).
As the exposure time used for all spectra is 120\,s, we can expect a point emission source to be blurred by $6.7^{\circ}$ azimuthally, which equates to a phase resolution of 0.019. Radial blurring will be the result of the finite spectral resolution of the data ($\sim 170\,{\rm km\,s}^{-1}$).

\begin{figure}
\centering
\includegraphics[width=1.03\columnwidth]{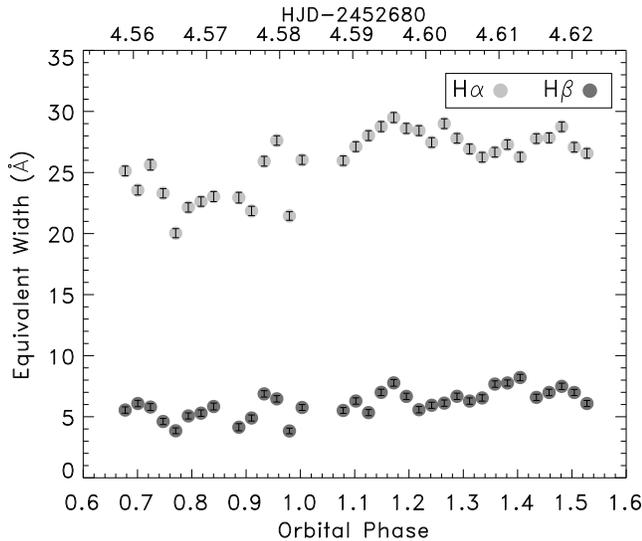}
\caption{Orbital variation of line equivalent width for H$\alpha$ and H$\beta$.}
\label{equiv}
\end{figure}

Modulation is apparent in the continuum level of the spectra, but it is not possible to distinguish real intrinsic modulation from that which can be attributed to variable slit losses. Hence, the examination of orbital variation in line flux is inappropriate. Instead orbital variations of equivalent width in H$\alpha$ and H$\beta$ are plotted (Figure \ref{equiv}). The underlying white dwarf absorption makes the calculation of true equivalent width somewhat problematic, and we did not attempt this for the higher order Balmer lines. Nonetheless the relative values for H$\alpha$ and H$\beta$ should be reliable. The figure suggests a weakening of H$\alpha$ from approximately phase 0.68 to phase 0.91. Without coverage over multiple orbits, it is uncertain whether this is a systematic variation or due to flickering. One of the underpinning assumptions of Doppler tomography is that line flux is constant over the orbit. To fulfill this we normalized the spectra for each map to constant integrated line flux, except where the effects of noise then produced significant artefacts in the maps, in H$\delta$, He\,{\sc ii}, Ca\,{\sc ii} and Fe\,{\sc ii}.

The S-wave in the trailed spectrograms maps into a concentrated region of emission in velocity space. We estimated the systemic velocity for the Doppler maps in two ways. The first was to make the sharpest possible maps, i.e. to use the value of $\gamma$ which minimised the width of a Gaussian fitted to the concentrated emission spot in the Doppler map. The average $\gamma$ value found by this method for the Balmer lines and He\,{\sc i}\,$\lambda5876$ is $29\,{\rm km\,s}^{-1}$, whilst the spread in values was $18$ to $46\,{\rm km\,s}^{-1}$. Fitting to the entire map as opposed to isolating just the emission spot yielded an average value of $56\,{\rm km\,s}^{-1}$, whilst a search for the $\gamma$ value which maximized the strength of the emission gave $41\,{\rm km\,s}^{-1}$. We adopted $\gamma=40\pm20\,{\rm km\,s}^{-1}$ for the maps. The maps produced using any of these differing $\gamma$ values are essentially indistinguishable.

\subsection{Gravitational redshift of the white dwarf} \label{grav}
\nocite{SCS97}{Sion} {et~al.} (1997) reported gravitational redshift of $58\pm33\,{\rm km\,s}^{-1}$ after estimating $\gamma=2\pm14\,{\rm km\,s}^{-1}$, from the radial velocity curve of \nocite{SV81}{Schoembs} \& {Vogt} (1981); \nocite{SV81}{Schoembs} \& {Vogt} (1981) however state that their $\gamma$ value is arbitrary and chosen so that the mean radial velocity is zero. The narrow absorption feature in the region of He\,{\sc i}\,$\lambda4472$ is Mg\,{\sc ii}\,$\lambda4481$ and originates in the white dwarf. In the average spectrum this feature has a shift of $78\pm8\,{\rm km\,s}^{-1}$. We can use the difference between the emission line systemic velocity, $40\pm20\,{\rm km\,s}^{-1}$, and the white dwarf Mg\,{\sc ii}\,$\lambda4481$ absorption line, to estimate the white dwarf's gravitational redshift. We obtain $v_{\rm grav}=38\pm21\,{\rm km\,s}^{-1}$. This is consistent with \nocite{SCS97}{Sion} {et~al.} (1997)'s result and is more secure since it is derived relative to simultaneously measured Balmer emission from the low-gravity environment of the outer disc.
Using this value of $v_{\rm grav}$, together with the mass-radius relation for a 20\,000\,K white dwarf with a carbon core \nocite{Wo95}({Wood} 1995), we find $M_{1}=0.71^{+0.18}_{-0.26}\,{\rm M_{\odot}}$, a white dwarf radius, $R_{1}=9.3^{+2.5}_{-2.6}\times10^{8}\,{\rm cm}$, and surface gravity, ${\rm log}g=8.04^{+0.38}_{-0.41}$.

\subsection{Source of the S-wave} \label{swave}
\nocite{vADG87}{van Amerongen} {et~al.} (1987) provide a photometric ephemeris of maximum light from the hotspot; the phasing of conjunction is unknown. Hence there are two possible scenarios for the origin of the S-wave: the stream-disc impact region; or the irradiated face of the donor. Neither hypothesis can be immediately ruled out. By varying the system parameters for the map of H$\gamma$, the likelihood of each was investigated. Only in the most extreme case that $M_{\rm WD}=1.44\,{\rm M_{\odot}}$, i.e. the Chandrasekhar mass, could emission correspond to the inner face of the donor. \nocite{S99}{Sion} (1999) finds the mean mass of 76 CV white dwarfs to be $0.86\,{\rm M_{\odot}}$, while \nocite{W90}{Webbink} (1990) finds a geometric mean primary mass for systems below the period gap to be $0.61\,{\rm M_{\odot}}$. A single object of course does not have to adhere to such averages, but $\sim1.4\,{\rm M_{\odot}}$ would be atypically large. Furthermore, a larger inclination ($i=75^{\circ}$) than estimated for VW\,Hyi and an ultra-low mass ratio ($0.04$) would be required. On the other hand, as Figure \ref{trail} shows, the S-wave emission can be naturally attributed to the stream-disc impact region using parameters consistent with other studies of VW\,Hyi. The coincidence, to within 0.01 in phase, of the emission spot with the photometric hotspot ephemeris (i.e. when no phase offset is applied) is also persuasive.

It is not possible to locate the position of the hotspot exactly. It would be expected to lie somewhere in the region between the ballistic stream trajectory and its Kepler shadow. A mixing of the stream and disc material and their respective velocities is expected as the material is compressed between two shock surfaces where the in-falling accretion stream impacts the disc edge and merges with the Keplerian flow \nocite{MHS90}({Marsh} {et~al.} 1990). An orbital phase offset of $0.15\pm0.04$ (where the error comes from extrapolating the epochs) with respect to the photometric ephemeris of \nocite{vADG87}{van Amerongen} {et~al.} (1987) was required to place the hotspot in H$\gamma$ on the ballistic stream trajectory close to the intersection with the tidal radius. This phase offset is comparable with that between $\phi_{0}$ and the hotspot orbital hump for the similar, but eclipsing, system, Z\,Cha, of 0.18 \nocite{W74}({Warner} 1974).

\subsection{Balmer emission}

Figure \ref{trail} shows the Doppler maps in the Balmer lines, constructed using the primary mass found in Section \ref{grav} combined with $q$ calculated from the superhump excess, and using the phase offset found above. In order to suppress high-frequency noise, a Gaussian cutoff is included in the Fourier filter that is applied to the spectra prior to back-projection. This Gaussian is chosen so that a suitable trade-off between resolution and noise is obtained. Here, the FWHM of the Gaussian window function is 10, and so there is minimal reduction in noise to preserve resolution.

There is a potential problem with contamination by the white dwarf absorption features, as they would be expected to have a different (unknown) phasing to that of the (hotspot) emission. To account for this, a model white dwarf spectrum was constructed, based on the properties found for the white dwarf in VW\,Hyi by \nocite{GSC04}{Godon} {et~al.} (2004) from UV observations. This model was scaled in flux to match our data at the blue end (Figure \ref{average}), and then fitted to the absorption wings of the Balmer line profiles in the individual spectra by allowing shifts in wavelength. The fitted white dwarf model spectrum was then subtracted. H$\gamma$ has the cleanest absorption wings in our data and so these were used in the fitting, although using the wings of H$\delta$ produced very similar results. The wavelength shifts obtained in this way were not inconsistent with the predicted motion of the white dwarf, if the parameters we have used in producing the Doppler maps are assumed \nocite{Smith}({Smith} 2006). It was found that most of the structure seen in the Doppler maps was unaffected by this procedure, the hotspot emission particularly so. What did change was the appearance of a confined central emission feature in the region of the white dwarf. The central regions of the lines corresponding to this low velocity emission will be where the effect of the correction is greatest owing to the deep absorption cores of the white dwarf spectrum. This feature must, therefore, be considered an artefact due to inaccuracies in the fitting procedure and uncertainties in the appropriateness of the parameters used in constructing the model spectrum. We nevertheless conclude that the effect on the Doppler maps of the white dwarf absorption is very minimal and does not need to be considered further.

The maps reveal how the systematic variation through lines of the Balmer series seen in the trailed spectrograms relates to distribution in velocity space, from a smoother distribution over the disc in H$\alpha$, to a concentrated hotspot in H$\delta$.

\begin{figure}
\centering
\includegraphics[width=\columnwidth]{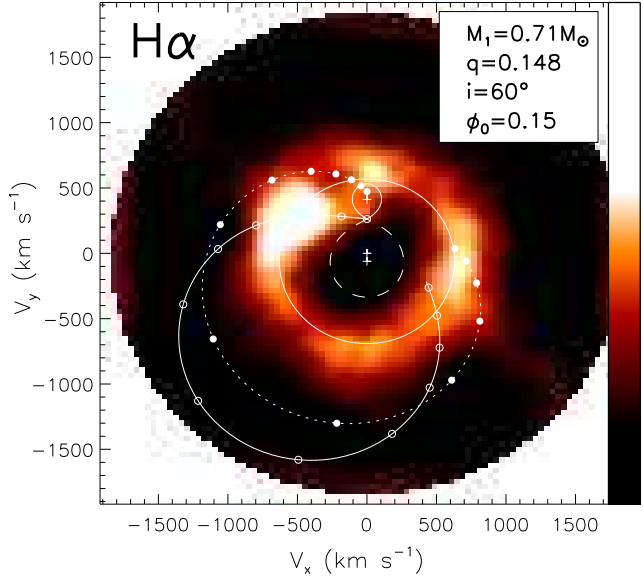}
\caption{Doppler map of H$\alpha$ emission using the maximum entropy method. Symbols are as in Figure \ref{trail}.}
\label{mem}
\end{figure}

The overplotted sine waves in the trailed spectrum of H$\alpha$ in Figure \ref{trail} indicated that the variable double-peaked emission may be correlated. A possible source for this emission is a secondary region of enhanced emission seen in Figure \ref{mem}. Here the maximum entropy method (MEM) is used (discussed in detail in \nocite{MH88}{Marsh} \& {Horne} (1988)) as opposed to FFBP. Both methods produce qualitatively very similar maps, depending somewhat on the value of the controlling parameter used (FWHM of noise filter in the case of FFBP, and $\chi^2$ in the case of MEM), but MEM appears to better detect detail such as a secondary emission region as is the case here.

\subsection{He \,{\sc i} and He \,{\sc ii} emission} \label{hetxt}

\begin{figure*}
\centering
\includegraphics[height=9.7cm]{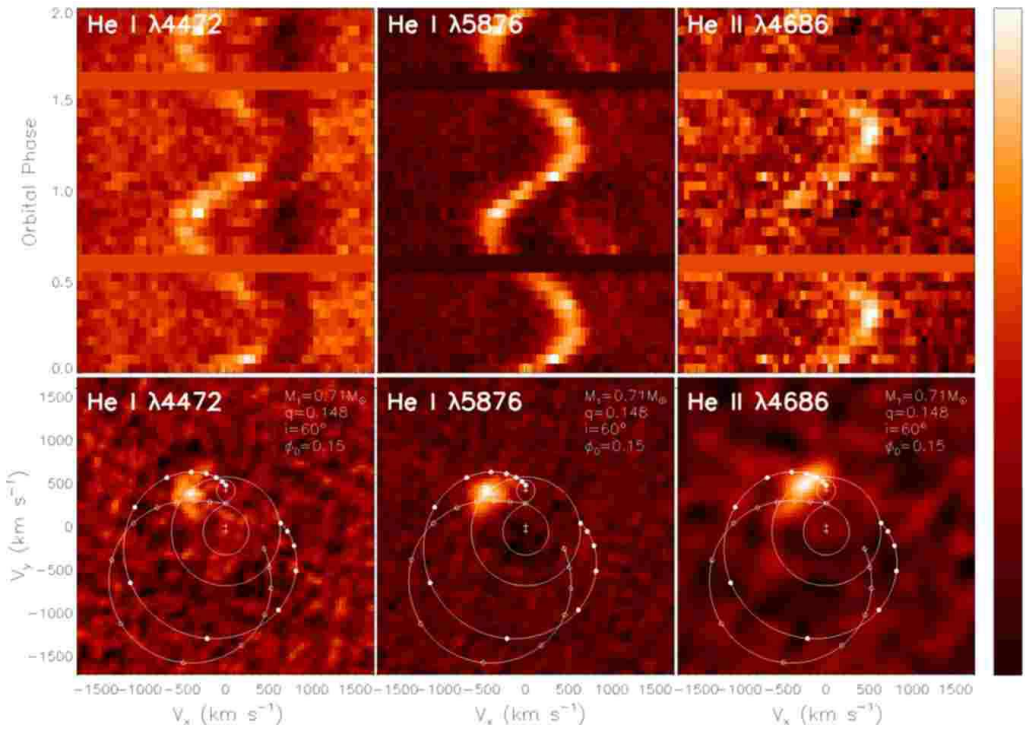}
\caption{Top panels left to right show the trailed spectra of He\,{\sc i}\,$\lambda4472$, He\,{\sc i}\,$\lambda5876$ and He\,{\sc ii}\,$\lambda4686$ respectively, and below the corresponding Doppler maps. Symbols are as in Figure \ref{trail}.}
\label{he}
\end{figure*}

Figure \ref{he} shows the trailed spectra and corresponding Doppler maps for He\,{\sc i}\,$\lambda5876$ and He\,{\sc i}\,$\lambda4472$. The Mg\,{\sc ii}\,$\lambda4481$ absorption feature from the white dwarf creates a dark ring artefact in the He\,{\sc i}\,$\lambda4472$ map. In both lines there is little or no evidence for a disc component in emission, while a strong S-wave is exhibited, again originating in the stream-disc impact region.  The He\,{\sc i}\,$\lambda5016$ and He\,{\sc i}\,$\lambda4922$ lines are possibly contaminated by blending with Fe\,{\sc ii}\,$\lambda5018$ and Fe\,{\sc ii}\,$\lambda4924$ respectively, and Doppler maps for these lines show a very much less distinct hotspot.
Of particular note is the presence of He\,{\sc ii}\,$\lambda4686$ in the spectra. This also demonstrates S-wave motion, with the emission, in contrast to that of the Balmer lines, strongest in the transition from blue to red. This S-wave also maps to approximately this same hotspot region (Figure \ref{he}) indicating a high temperature at the hotspot. Because He\,{\sc ii}\,$\lambda4686$ is a significantly weaker feature in the spectra, noise posed a greater problem in the production of the Doppler map. Here a Gaussian window function with a FWHM of 0.3 was used with the inevitable loss of resolution, but still the hotspot remains clear.

That He\,{\sc ii} is detected and seen to emanate from the hotspot region in a quiescent DN is relatively rare. It was seen in U Gem \nocite{MHS90,G01}({Marsh} {et~al.} 1990; {Groot} 2001), where the velocity of emitting material was found to lie intermediate between that of the accretion stream and the Keplerian disc at the stream-impact, and in the newly identified cataclysmic variable H$\alpha$0242-2802 \nocite{MH05}({Mason} \& {Howell} 2005).

\subsection{Emission in metal lines} \label{caiitxt}

\begin{figure}
\centering
\includegraphics[width=\columnwidth]{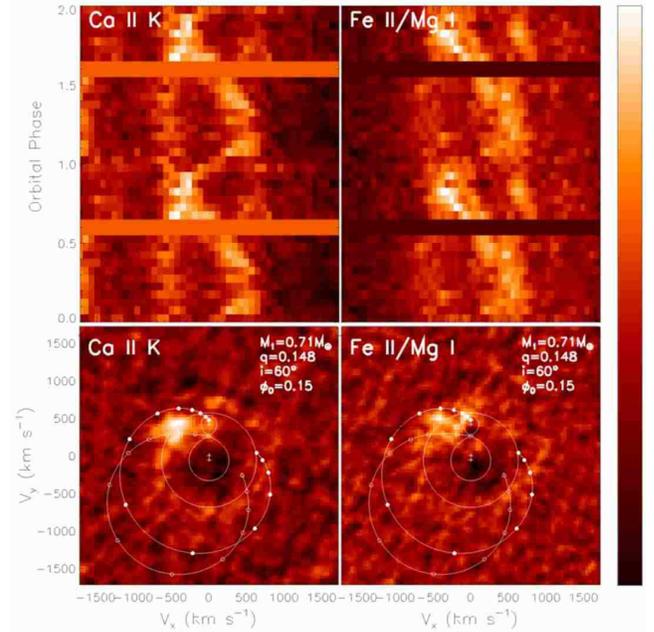}
\caption{Top left and top right panels show trailed spectra of Ca\,{\sc ii}\,K and Fe\,{\sc ii}\,$\lambda5169$ respectively, and below the corresponding Doppler maps. Symbols are as in Figure \ref{trail}.}
\label{caii}
\end{figure}

Figure \ref{caii} shows the corresponding trailed spectra and Doppler maps for Ca\,{\sc ii}\,K and Fe\,{\sc ii}\,$\lambda5169$ respectively. Fe\,{\sc ii}\,$\lambda5169$ may be blended with Mg\,{\sc i}, blurring the hotspot.
The Ca\,{\sc ii}\,K hotspot is clear. There is a suggestion of weak disc emission in both Ca\,{\sc ii} and Fe\,{\sc ii}.

\section{Interpretation}

\subsection{The Balmer Doppler maps}

The Doppler maps of the Balmer lines show a progression from hotspot emission superimposed on disc emission in H$\alpha$ to a relatively more dominant hotspot in H$\beta$, to solely hotspot emission in H$\delta$. This may be interpreted as the hotspot being optically thick in H$\alpha$ and less so in H$\beta$, so that in H$\alpha$ the hotspot is weaker relative to the optically thin disc emission.
This behaviour is seen in the Doppler maps of other quiescent dwarf novae \nocite{SMH00,RAH01}({Skidmore} {et~al.} 2000; {Rolfe}, {Abbott} \& {Haswell} 2001), and in the quiescent soft X-ray transient A0620-00 \nocite{SHC04}({Shahbaz} {et~al.} 2004). In the latter system, a significant modulation of H$\alpha$ with orbital phase is observed, perhaps as a result of a changing perspective of an optically thick, elongated hotspot.

In Figure \ref{mem}, a second region of enhanced emission in H$\alpha$ is seen. The limited phase coverage of our observations limits confidence in the reality of this feature. The dark band across the maximum-entropy map running from bottom left to top right for example, corresponds to the phase gap in the observations, which could also influence the appearance of this second enhanced emission feature. We note that the phase at which the presence of a putative secondary S-wave is most convincing corresponds to when H$\alpha$ emission was strongest (Figure \ref{equiv}). If real, then it would indeed be interesting, and we are without an explanation for its presence. It does not fall where an overflowing accretion stream would be expected to re-impact the accretion disc, cf. \nocite{KSH01}{Kunze} {et~al.} (2001) and their figure 13.
\nocite{L93}{Livio} (1993, and references therein) discuss secondary interactions between the accretion stream and disc, and observation of multiple enhanced regions of emission can be found in \nocite{MaH90,N98,N02,RGM05}{Marsh} \& {Horne} (1990); {Neustroev} (1998, 2002); {Roelofs} {et~al.} (2005).

\subsection{Ionization structure of the accretion stream-disc impact region} \label{ionization}

\begin{figure}
\centering
\includegraphics[width=\columnwidth]{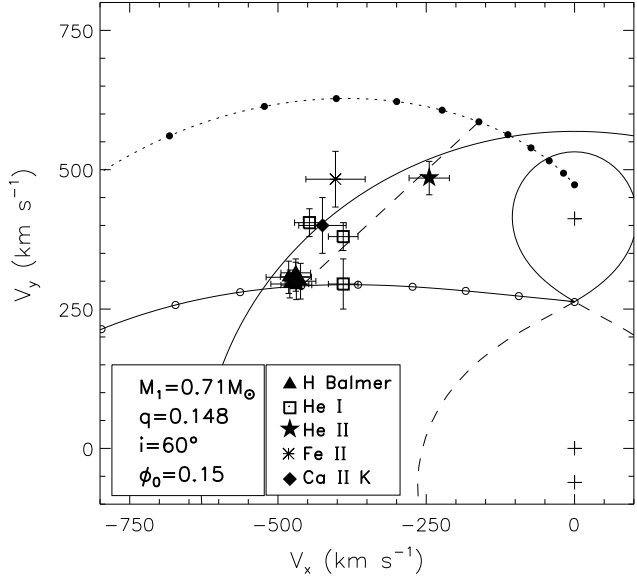}
\caption{Detail of hotspot region in velocity space with the centroid locations of the hotspots in the emission lines shown. Symbols used are indicated in the legend. Additional symbols used are as in Figure \ref{trail}. The dashed line connects the ballistic stream and Keplerian disc velocities at a radius of $0.5\,R_{\rm L1}$ for the system parameters assumed.}
\label{hotspot}
\end{figure}

\nocite{PBH87}{Pringle} {et~al.} (1987) estimated the colour temperature of the orbital hump due to the hotspot in VW\,Hyi to be $11\,000\pm2000\,{\rm K}$, which is typical of the blackbody temperature of the hotspot in DNe. Our observations show that temperatures in the stream-disc interaction region must be sufficient to ionize helium. We show the locations of the centroid of concentrated hotspot emission (with $\phi_{0}=0.15$ assumed) in Figure \ref{hotspot}. As the true phasing of the system is unknown, these positions may be rotated around the centre of mass of the system, but their relative positions will remain as shown. The hotspots in the Balmer and He\,{\sc i} lines lie downstream of the hotspot in He\,{\sc ii}\,$\lambda4686$. This situation would be expected if the temperature of the hotspot region is such that neutral emission is suppressed by ionization at the impact, and that recombination can only take place once the stream material mixing with the disc material has moved downstream of the initial shock interaction and cooled sufficiently. As the first ionization potential for Ca is less than that of He, so we can also expect Ca\,{\sc ii}\,K hotspot emission to be seen a lower temperatures downstream of the He\,{\sc ii} emission, just as Figure \ref{hotspot} indicates.
From eclipse measurements in WZ\,Sge, \nocite{SR98}{Spruit} \& {Rutten} (1998) found the hotspot in the optical continuum and in H$\alpha$ to be separated in phase by 0.015, with the continuum leading H$\alpha$. In IY\,UMa, emission in the Balmer lines is found to be weaker in the disc regions around the stream-disc impact \nocite{RHA05}({Rolfe} {et~al.} 2005), indicating that Balmer emission is indeed suppressed at the high temperatures of the hotspot. We similarly see a suppression of disc H$\alpha$ emission along the Kepler shadow of the stream (Figures \ref{trail} and \ref{mem}). \nocite{SR98}{Spruit} \& {Rutten} (1998) invoke a small, fully ionized `Str\"omgren sphere' around the hotspot to explain their observations of WZ\,Sge: intense UV emission causing a region of reduced $H\alpha$ brightness in the disc bordering the impact point.

However, the same physical position can map to differing velocity locations, and it is also possible that the hotspot emission in the different lines emanates from the same physical point but that the dynamics of emitting material differs. Another possible interpretation of Figure \ref{hotspot} therefore, is reached by noting that the He\,{\sc ii} hotspot is closer to the Kepler shadow than the Ca\,{\sc ii}, He\,{\sc i} and Balmer hotspots. This might indicate that He\,{\sc ii} emission may arise from material with predominately disc-like velocities, the emission perhaps taking place after a greater degree mixing with the disc material, while the Balmer emission is dominated by material in the ballistic stream.
Note $\phi_{0}=0.15$ was found by placing the H$\gamma$ hotspot on the stream trajectory. If we had instead placed the He\,{\sc ii} hotspot on the stream trajectory (effectively rotating the points in Figure \ref{hotspot} about the centre of mass, while leaving the stream and Roche lobe loci fixed) then the hotspot emission in the other species would fall below the stream trajectory in the map. In eclipsing systems where the phasing is known \nocite{RHA05}(e.g. IY\,UMa, {Rolfe} {et~al.} 2005) the Balmer and He\,{\sc i} hotspots occur between the stream trajectory and its Kepler shadow as they do with our choice of $\phi_{0}$.

The differing velocities of the hotspot in the various spectral lines shows this localized emission does emanate from the stream-disc impact. An origin on the mass donor star would require all species to have velocities consistent with the donor star. Figure \ref{hotspot} shows, however, the spread in velocities is bigger than the velocity image of the donor Roche lobe.

Emission in He\,{\sc ii}, in contrast to He\,{\sc i}, shows a strong phase dependence (Figure \ref{he}) (the apparent phase dependence in  He\,{\sc i}\,$\lambda4472$ we attribute to the Mg\,{\sc ii}\,$\lambda4481$ absorption). He\,{\sc ii} emission requires high temperatures which, in quiescence, will likely only be reached at the initial shock-heated stream-disc impact point, hence He\,{\sc ii} emission is expected to be highly localized. \nocite{SPR03}{Steeghs} {et~al.} (2003) discovered an unusual feature in their orbital lightcurve of IY\,UMa: a rise occurring between the end of white dwarf ingress and hotpot ingress. This is best explained by the brief view into a shock-heated cavity with a narrow opening angle, caused by the impact of the accretion stream. A similar cavity, though wider (as VW\,Hyi has a lower inclination), could cause the phase dependence of He\,{\sc ii}. The emission might be expected to be seen at phases where the hotspot faces us, reaching a maximum at the point where we are looking into the cavity near maximum redshift, as our data shows. He\,{\sc i} emission, on the other hand, probably comes from a larger physical volume at lower temperatures, and is hence visible throughout the orbit.

\subsection{System parameters revisited}

\begin{figure}
\centering
\includegraphics[width=\columnwidth]{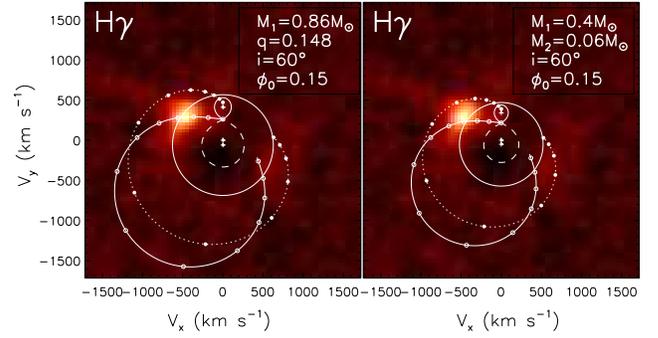}
\caption{Doppler maps of H$\gamma$ showing the location of the hotspot using various parameters in the literature, and assuming $\phi_{0}=0.15$, as shown in the legend of each panel. Again, symbols are as in Figure \ref{trail}. The Doppler map in the left panel uses the value for $M_{1}$ given by Sion et al. (1997), where $q=0.148$ is assumed, while that in the right panel uses the values of Mennickent et al. (2004).}
\label{parameters}
\end{figure}

The radius of the accretion disc is unknown. Because of this uncertainty it was not possible to discriminate decisively between the various system parameters put forward in the literature, and all, including the $M_{1}$ value we derive in Section \ref{grav}, appear to be consistent with the hypothesis of the hotspot as being the source of the S-wave observed (Figure \ref{parameters}). From Figure \ref{hotspot} it might appear that the hotspot in several lines falls outside the tidal radius (the disc is inverted in velocity space).
Under the assumption of Keplerian disc velocities, the requirement for the position of the H$\gamma$ hotspot to be within the tidal radius places some tentative limits on the  system parameters for VW\,Hyi. If $q=0.148$ is assumed, then $M_{1}\sin^{3}\,i<0.39\pm0.05\,{\rm M_{\odot}}$ is required.
However, we may expect the velocities in the outer disc to be $\sim10\,\%$ sub-Keplerian for a system of this mass-ratio (Truss, private communication). This, coupled with the blurring in the Doppler maps, removes the apparent conflict between the hotspot positions and tidal radius in Figure \ref{hotspot}.
We note, though, that the disc is expected to reside well within $R_{\rm tides}$ when in a quiescent state.

\section{Summary}

The following summarises our main findings:

\begin{itemize}

\item We observed VW\,Hyi in quiescence. Slit losses mean $V<14.5$.
\item There is strong S-wave emission in the cores of the Balmer series and in He\,{\sc i}, He\,{\sc ii} and Ca\,{\sc ii}. This S-wave arises in the stream-disc impact region.
\item We favour a phase offset of $\phi_{0}=0.15\pm0.04$ between the photometric ephemeris and inferior conjunction of the mass donor.
\item A progression in the appearance of the Doppler maps, from hotspot emission superimposed on disc emission in H$\alpha$, to relatively more prominent hotspot emission in the higher order Balmer lines, is seen. This may be interpreted as the hotspot being optically thick in H$\alpha$ and less so in H$\beta$, so that in H$\alpha$ the hotspot is weaker relative to the optically thin disc emission.
\item The stream-disc interaction region shows an ionization structure in velocity space. One interpretation is that the Balmer and He\,{\sc i} hotspots lie downstream of the He\,{\sc ii} hotspot. This may be expected if immediate post-shock temperatures are such that H and He are ionized and recombination can only occur once material has cooled downstream of the impact point.
\item He\,{\sc ii} emission shows a strong phase dependence, which may be the result of restricted viewing of a localized emission site in a shock-heated cavity at the stream-disc impact site.
\item A secondary region of enhanced emission is detected in the Doppler map of H$\alpha$ produced by the maximum entropy method, though it is unclear how this might arise.
\item We detect broad Balmer absorption and narrow Mg\,{\sc ii}\,$\lambda4481$ absorption from the white dwarf. We use this Mg\,{\sc ii}\,$\lambda4481$ feature to determine the gravitational redshift, $v_{\rm grav}=38\pm21\,{\rm km\,s}^{-1}$.
\item Our preferred system parameters are $M_{1}=0.71^{+0.18}_{-0.26}\,{\rm M_{\odot}}$ (from our $v_{\rm grav}$) and $q=0.148\pm0.004$ from the superhump period excess \nocite{P98}({Patterson} 1998). This implies $M_{2}=0.11\pm0.03\,{\rm M_{\odot}}$. Hence the donor star would not be a brown dwarf.
\end{itemize}
Though He\,{\sc ii} in the hotspot of quiescent DN has been rarely detected before now, its presence may actually be a widespread phenomenon. The S/N of existing observations may simply not be sufficiently high to detect it. Our results show that modest amounts of 4\,m telescope time (less than 2 hours in this case) can resolve uncertainties regarding the fundamental parameters of CVs.

\section{Acknowledgements}

Many thanks go to Daniel Rolfe for his kind use of, and great help with, his {\sc IDL} routines; to Andy Norton and an anonymous referee for reading of drafts and helpful suggestions which led to improvement of the paper; to Matt Burleigh for providing us with a model white dwarf spectrum; to Michael Truss and Edward Sion for helpful discussion; and to Tom Marsh for use of his {\sc MOLLY} and {\sc DOPPLER} spectral analysis software and for suggestions regarding the use of MEM. Tim Abbott deserves particular thanks for help with technical difficulties, without which the observations presented here would not have been possible. This work has made use of the NASA Astrophysics Data System Abstract Service.
AJS was supported by a PPARC studentship. RIH was partially supported by
NASA through Hubble Fellowship grant \#HF-01150.01-A awarded by STScI and
through \textit{HST} grant GO\,9398.

\label{lastpage}

\end{document}